# Detection of pure epileptic transitory among gamma oscillatory activities in MEG signals by SVD, embedded on a partial dynamic reconfiguration

Nawel Jmail* (1), Tarek Frikha** , Abir Hadriche***,  Christian-G Bénar****.

*Sfax University, MIRACL, Sfax, Tunisia.
**Sfax University, CES, Sfax, Tunisia
***Gabes University, REGIM, Sfax, Tunisia.
**** Aix-Marseille University, Faculty of Medicine, INSERM, Institut de Neurosciences des Systèmes, UMR,1106 Marseille, France.

(1) naweljmail@yahoo.fr

**Abstract:**

In this work we used the SVD to reconstruct pure epileptic transitory activities for the definition of the accurate sources responsible of the excessive discharge implied by the transitory hallmark activities in pharmaco resistant epilepsy. We applied firstly an automatic detection of the local peaks of the transitory activities through a thresholding the energy distrubition, then we performed the SVD on the detectable transitory (300 ms window) simulated events to recover only the non contaminated transitory activities by gamma oscillations, we calculate the precision of the SVD to evaluate the robustness of this technique in separation between transitory and gamma oscillations. In second phase we applied the SVD on small window of transitory activities from real MEG signal, we obtained dipolar topographic map for the pure transitory activities, however the time excitation for the recover of pure epileptic transitory activities among the gamma oscillations were very heavy and hence we propose to integrate the SVD on a  dynamic partial  reconfiguration. This integration were very useful, in fact we did obtained about …. Times faster in the  SVD execution.

## I. Introduction:

Now days, the diagnosis of neurologic disease is oriented to the exploitation of the magnetoencephalography MEG technique thanks to it's major advantages [1]. In fact, the MEG recording allows an important time precision and also space one, and is more sensitive than the EEG for activities with restricted expended

[2] .The space precision of this registration technique is obtained through the large number of captors which can reach 248 captors. Hence, and despite it s cost, the neurologist and the biomedical researches are using this technique as a complementary or even primordial way to diagnosis the epilepsy and specially the pharmaco resistant, and to overcome the invasive technique of registration (the intracerbral EEG) [3]. The pharmaco resistant epilepsy needs a high space precision for the detection of the epileptogenic zone EZ (Jmail et al. 2016; Wendling et al. 2009), which would be in next level restricted by a surgical intervention. There for The neurologist rely on the networks connectivity of the MEG hallmark as the gamma oscillations [6] and the transitory activities non contaminated [4] to define the accurate EZ [7]. However, these activities are mutually contaminated due to the overlap in the time frequency plane [8], which would affect the responsible sources of the excessive discharges and the build up seizure. In previous work, we proposed and evaluated the performances of several filtering techniques in separation between transient and oscillatory activities (matching pursuit MP, finite impulse response filter FIR and stationary wavelet transform SWT) [9], we studied also the singular value decomposition SVD to despikify the IEEG signals (retrieving the transitory activities from the gamma oscillations) to predict the time and space build up seizure [10], [11]. Thus, we suggest, in this work to apply the SVD for the reconstruction of pure epileptic transitory activities among the MEG signals. The SVD technique is a decomposition step, based on the projection on a suited basis, followed by a thresholding step (applied on the results component responsible of the transient activities reconstruction) and a recover phase using only the selected components. A simple restriction between the constructed transitory activities among the original signal will result on the oscillatory activities. In this work, we started by simulating a signal with 5 channels that depicts an overlap between the transitory activities and the gamma oscillations, then we reproduce the pure transitory activities by SVD and we calculate the goodness of fit of the reconstruction depending on the gamma oscillation frequency range and the signal to noise ratio. For real MEG signal, we evaluate the robustness of recognition of pure epileptic spikes by SVD, we proposed to compare the topography

map of pure transitory activity and a mixture of gamma and transitory activities. In a second time, we will apply the same preprocessing scheme (high and low thresholding on the energy distribution to detect the local peaks) we expend the window to 300 ms around the peak to achieve an automatic detection of the MEG transitory activities, finally we will measure the true and false positive value and precision of the epileptic transient automatic detection. However, the application of the SVD for a large number of captors and a big data set is heavy in computation. Here we suggest to embed the SVD on a partial dynamic architecture. In fact, the integration of pre processing routines in the biomedical field is highly suggested to reduce the time consumption and to propose new neurofeedback systems to help the diagnosis of neurologist in real time. The integration of preprocessing algorithms rely on the study of the coherent hardware that allow the best match of the proposed software……………………………. ……………………………………. ………………………………..

This work is composed of three sections, the first one to describe the signals and methods (the detection, separation of the transitory activities by SVD and finally, the implementation of the SVD on an intelligent architecture) the second section is dedicated for the discussion and the last one for the conclusion and further perspectives.

## I. Materials and Methods
### I.1. Materials

#### I.1.1. Simulations

We simulate a dataset inspired from our MEEG acquisition, the sampling frequency was set to 1024 Hz. We produce the mixture between the transitory and gamma oscillatory activity through the translation of the gamma oscillations with equal steps across the transitory window, hence we obtained transitory and oscillations separated in the time overlapped and fully overlapped [8]. We added a noise with a physiologically plausible $1/f$ spectrum, generated from a neural mass model [4], and we vary the signal to noise ratio : SNR for three value -5, 15 and 20 dB , for each SNR value we made 100 realizations (300 in total). We simulate five channels each one depict the mixture of the transitory and oscillatory activities for a frequency range 45, 55, 65, 75, and 85 . We will study the effect of the SNR and the frequency range on the reconstruction of the pure epileptic transitory activities by the SVD technique.

#### I.1.2. Real Signal

Our real signal explored in this work was a MEG signal for a pharmaco resistant subject with Left occipito temporal cortectomy epilepsy. The acquisition and pre processing steps were applied in the Clinical Neurophysiology Department of La Timone hospital in Marseille, France. The MEG signal was recorded on a 151 gradiometers system (CTF Systems Inc., Port Coquitlam, Canada). The patient was awake with eyes closed, without activation procedure and movement. The sampling frequency was set to 1025 Hz, 20 epochs of 5s were recorded, and triggered visually on spike by the expert professor Martine Gavaret. The head position was digitized with the help of 3 coils placed on the 127 head in the start and finish of each run stored. A change of the head position above 5 mm necessitate waiving the actual run as in [10]. The patient studied in this paper is features by an abundant transitory activities and diagnostic as symptomatic epilepsy with a Left occipito temporal cortectomy localized in the Left occipito temporal junction.

Patients signed informed consent, and the study was approved by the Institutional Review board (IRB00003888) of INSERM (IORG0003254, FWA00005831).

The simulated signals and all signal processing was done using the Matlab software (Mathworks,Natick, 99 MA),and the EEGlab toolbox [12].

….VHDL

**I.2.Methods**

**A/ Detection of transitory shapes**

To detect, the transitory activities among the simulated and real MEG signal we proceed as in [11] , we applied , for each channel separately (since the shape of the transient activities can change from a captor to another ) a high and low thresholding step applied on the amplitude distribution Qp of 0.5, 0.75 and 0.25 percentile. We imposed 10 ms distance between tow consecutive peaks as explained in the equation (1):

$$thr_h = Q_{0.5} + d(Q_{0.75} - Q_{0.25})$$
$$thr_l = Q_{0.5} - d(Q_{0.75} - Q_{0.25})$$

Hence, we obtained the local peaks of the transitory activities among simulated and real data, we create transitory epochs around the peaks with respect to the distance constraint between two consecutive activities (set to 10ms).

**B/ SVD**

After, the automatic detection of the transitory activities applied in last section, we segment our data base (for simulated and real MEG signal), in events lasting 300 ms around the peaks detected previously. Then we performed the SVD on these epochs and for each channel consecutively [13]. In fact, the singular value decomposition SVD is a pre processing technique that allow the construction of a new signal through a number of component [14] and since our mission is the separation between epileptic transients and gamma oscillations the SVD results on several components, thus we applied a thresholding steps to select the transitory components from the oscillatory one. The singular value decomposition of a given signal X is σ that verifies the equation (2) below:

$X*V = \sigma U$ and $X*U = \sigma V,$

Where X is a matrix with (m, n) dimension, U is a non zero n left singular vector and V is a non zero m right singular vector.
Hence, the singular value decomposition of X is the reduction of X to a bi-diagonal matrix obtained through the product between 2 orthogonal columns U and V and $\Sigma$ a diagonal matrix with a nonnegative value as defined in the equation (3)
$X = U\Sigma V$

We applied the SVD on a window of 300 ms of the detectible transient events by the quartile methods (previous section) for each channel (simulated and real data). We create a new basis by computing the first 3 sum product of the time U(t) and space V(s) components as in equation (4):

$$T(t) = \sum_{1}^{3} \Sigma_i u_i v_i$$

The last step of the reconstruction of the transitory activities is the projection of the original signal (simulated and real MEG signal on the model constructed by the first three components of the SVD technique) [15].

**C/ GOF**

The Goodness of fit GOF is a measure that reflects the robustness of the transitory activities reconstructed by the SVD filtering technique among the original signal which contain a mixture of gamma oscillations and transitory activities [9]. The GOF is the ratio between the energy of the pure transitory activities obtained by the SVD per the energy of the original signal that depicts both activities (gamma oscillations and transitory activities) as defined in the equation (5).

$$GOF = \frac{\sum(x(t) - \widehat{x(t)})^2}{\sum x(t)^2}$$

We computed the GOF for the 5 range of gamma oscillations 45, 55, 65, 75 and 85 and for the 3 value of SNR: -5 10 20. In fact the GOF results will define the robustness of the transitory activities separation from the gamma oscillations according to the SNR and frequency range that could vary from a data set to another. We will compare the performance of the SVD in the reconstruction of the pure epileptic transitory activities for simulated data for different frequency and SNR rate to define the robustness of the SVD versus the noise and the rapid discharge.

**D/ Clustering**

The epileptic transitory activities selected for the first run of MEG signal by the expert MG reach 50 events, however theses epochs did vary in shape from one channel to another. These activities could present a hyper polarization followed by depolarization or versa and could be illustrated by one phase (hyperpolarization or depolarization) followed by a discharge or overcharge. Hence we applied a clustering step to classify the transitory activities that share the same spatio temporal shape and imply the same active sources responsible for these excessive discharges. We proceed firstly by a temporal translation to align the transitory peaks then we used the k means algorithm to classify these activities [16]. The expert MG imposed the number of group to 2 clusters [17]. We used the *erpimage* function of the EEGlab toolbox to define the accurate events of each cluster [12].

**E) Precision**

First, the expert neurologist professor Martine Gavaret made a visual marking of the epileptic transitory activities on real MEG signal, then we performed a comparison between the results of the expert selection and the automatic detection and reconstruction of the pure epileptic spikes by the SVD technique [18]. To evaluate the performance of the SVD automated detector versus the neurologist selection, we calculated the precision P see equation (6)

$$P = \frac{TP}{TP + FP}$$

Where, TP is the true positive, FP is the false positive, and P is the precision measure that define the deviation between true and false values [18]. In order, to push ahead the robustness of the SVD automatic separator of the pure epileptic transitory activities, we lead the expert detection as in [9] guided by the automatic detection results obtained, hence another human detection were reapplied inspired by the automatic detection [19].

**F/ Topography**

To evaluate the robustness of the SVD reconstruction of the pure epileptic transitory activities on real MEG signal, we average all the selected transitory activities that belong to each cluster then we compared the topography of the original signal (selected transitory that present a mixture of transitory and gamma oscillations) versus the detectable transitory activities reconstructed by the SVD technique. These topography maps were calculated on the peaks of the transitory activities events. In fact the topography maps illustrate a 2 dimension representation of multi channel on scalp. A dipolar topography imply a cerebral activity, however a random activity could be the results of noise or induced a non cerebral activity [20].

## G/ The integration of the SVD technique

## II. Results and Discussion

In figure 1 we depict one realization of the simulated data a mixture, between transitory activities with different level of overlap (the oscillatory activities is translated in equal steps through the transitory window). We obtained 5 noisy channels each one presents the transitory and the gamma oscillations in this order: gamma oscillations with 45 Hz, 55Hz, 65Hz, 75Hz and 85 Hz. In the figure1 we illustrate one realization for SNR=10;

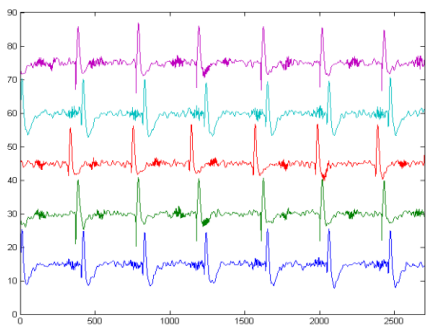

Fig1: One realization of simulated data, five channels that reproduce an overlap between transient and gamma oscillations, each channel deal with a range of frequency 45,55,65,75 and 85 Hz.

In figure 2 we present the detection of the transitory activities by the definition of the local peaks.

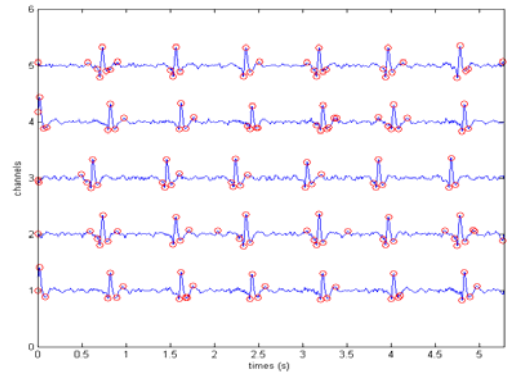

Fig 2: Detection of the transitory activities : in blue is the simulated data a mixture between transient and gamma oscillations , the red circle is the low and high thresholding results to detect the local peaks of the transient activities.

In figure 3 we present the reconstruction of the transitory activities among the gamma oscillations of 85 Hz by the SVD technique using the first three components.

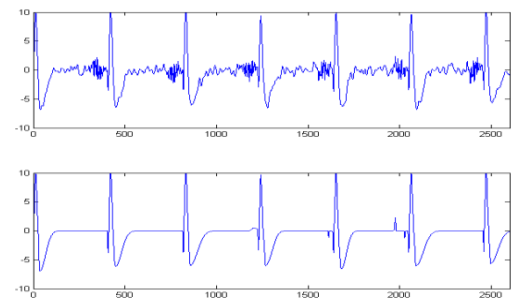

Fig3: line one channel 5 of simulated data depicting the mixture of gamma oscillations of 85 Hz and transitory activities, line 2 reconstruction of pure epileptic transitory activities among the gamma oscillations by SVD.

In figure 4, we depict the robustness of the transitory activities reconstruction by the SVD for the 5 range of gamma oscillations (45, 55, 65, 75 and 85) and for the three range of SNR [-5 10 20]dB.
 The GOF is computed between the original transitory simulated versus the reconstructed transitory detect by thresholding and reconstructed by the 3 first components of SVD. We notice that the GOF is ameliorated with the increase of frequency for a low gamma the GOF is about 80 % of fit and could reach the 90% of resemblance for 85 Hz oscillations, these results was also validated in previous work for several filtering technique as the MP, FIR ans the SWT (jmail et al 2011, bénar 2009). Also the GOF results depend straightly of the level of noise in fact for low SNR the GOF wouldn't exceed the 60% of reconstruction however for a good SNR the GOF could reach almost the 93% of resemblance.

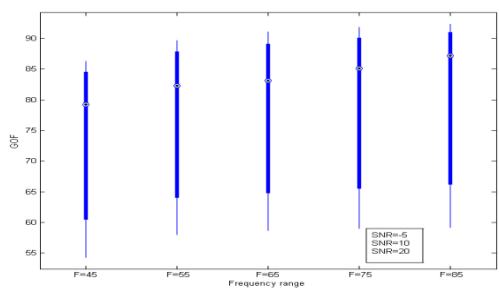

Fig4: GOF of the pure reconstructed transitory versus the simulated transitory mixed with the gamma oscillations for frequency range [45 85] and for SNR= -5 10 20 dB.

In figure 5, we illustrate our real MEG signal with 151 captors, the signal depicts three kind of activities oscillations, transitory activities and mixture of transitory and gamma oscillations (selected by an expert neurologist MG).

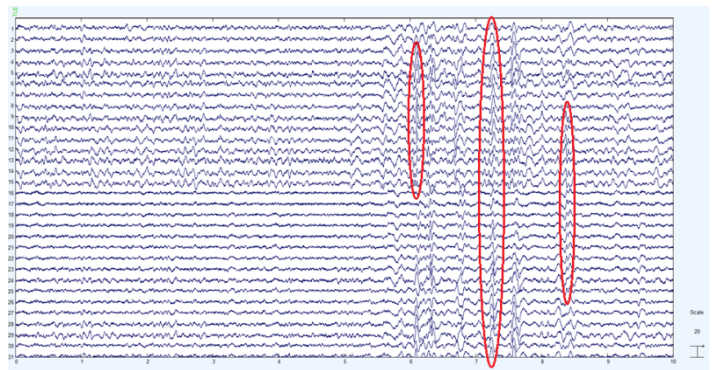

Fig5: MEG signal, only 31 captors are highlighted with an active appearance of mixed transitory activities, circle in red shows different shape of transitory activities.

In figure 5, we present the results of clustering the transitory activities among the real MEG signal, two groups of transitory activities are separated. The first group is composed by the transitory events from epochs 1 to 14 and the second cluster include the transitory events from epochs 24 to 34.

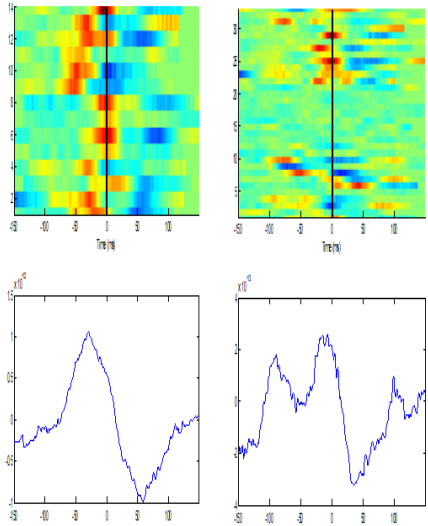

Fig5: Classification of MEG transitory activities on 2 groups line 1 the energy map of selected transitory for group 1and 2, line 2 the shape of the transitory clusters.

In figure 6 we depict the results of detection of local peaks of MEG transitory activities among a mixture of transients and gamma oscillations by high and low distribution thresholding.

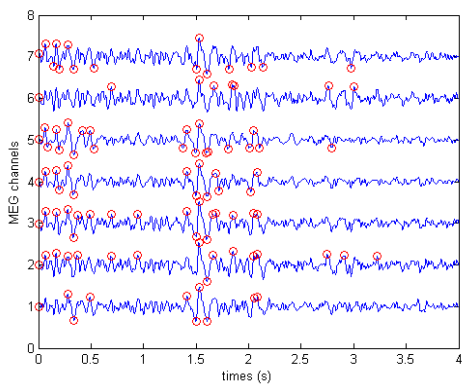

Fig 6: Detection of the MEG transitory shapes, the red circles detect the local peaks of each transitory activities.

In figure 7, we present , after the detection, the reconstruction by SVD of the pure epileptic transient activities by the three first components of the SVD technique. We notice that all the transitory activities were successfully recovered without no oscillatory activities: non contaminated epileptic spikes (transitory).

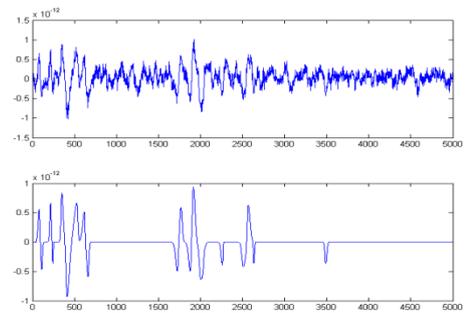

Fig 7: first line shows the Channel 25 a mixture of activities and line 2 presents the reconstruction of the real MEG pure transitory activities among the real MEG signal.

In figure 8, we depict the topography on the peaks of the transitory activity of the channel 25 (presented in figure 7) at sample number 1965 (which corresponds to the peaks of a transitory activity) before and after applying the detection and the reconstruction by SVD. It s obvious that pure epileptic transitory has a dipolar topography which reflects the presence of an accurate cerebral source responsible of the generation of this transitory however the topography of mixed transitory and oscillations shows much more random and complex activities that made the definition of responsible sources very difficult and hard to explain physiologically.

Hence the filtering technique applied by the SVD technique after the automatic detection lead to a better characterization of the responsible sources since the topography map of the reconstructed transitory activities illustrates only a dipolar activity and no further activation this will simplify the definition of the accurate epileptogenic zone EZ to be in next steps retrieved by a chirurgical intervention to stop or reduce the pharmaco resistant seizure.

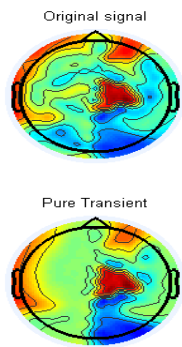

Fig 8: line one topography of original selected transitory, line 2 topography map of pure detected and reconstructed transitory activity by SVD. The second topography depicts a dipolar configuration which reflect a cerebral source.

In table 1, we collect the result of SVD robustness in detection and reconstruction (TP, FP and P) of transitory activities among the gamma oscillations from a real MEG signal. These results were obtained after the second human detection guided by the first results of automatic detection applied. We proceed on the three runs of MEG signal. These measures were computed for 4 channels (MLC14, MLC15, MLC33, MLC43) since they are featured by a large number of transient activities.

Table 1 : Precesion of automatic detection and reconstruction of the transient activities by SVD.

| Channels | TP | FP | P |
|----------|-----|----|------|
| MLC14 | 194 | 28 | 0.87 |
| MLC15 | 187 | 32 | 0.85 |
| MLC33 | 201 | 46 | 0.81 |
| MLC43 | 198 | 38 | 0.83 |

The precision of automatic detection and reconstruction of the transient activities vary from 81% to 87% (a high level of convenient automatic detection) . These results are very prometer and would led to a high automatic recognition of the epileptic transient by the SVD technique . These results will imply positively the definition of the epileptogenic zone EZ that should be restricted surgicaly for seizure free.

## III. Conclusion and Perspectives

The interpretation of the electrophysiological signals is a successful the way to define and diagnosis the accurate sources responsible of the cognitive and neurological lesion. In this work, we used the MEG signals to help neurologist in making decision in pharmaco resistant epileptic diagnosis. The MEG signal is featured by it s large captors which leads to a better time and space precision. In this work we applied an automatic detection of the local peaks of the transitory activities , then we applied the SVD technique to recover only the pure transitory activities among the gamma oscillations. We computed the GOF for different SNR and frequency to evaluate the performances of the SVD reconstruction of the non contaminated transitory activities. In fact , the precision measure prove the efficiency of the transitory activities reconstruction. Then, we proceed in the same way for the real MEG signal in detection and reconstruction of the pure epileptic transitory activities, and in order to evaluate the performance of the preprocessing scheme , we apply the topography map for original and reconstructed transitory activities; the topography map for the non contaminated transitory were dipolar which prove a cerebral source and explain the importance of the pure transitory activities as a hallmark for the definition of the epileptogenic zone.

However the computation of the SVD for a large number of captors and a big data set seems to be very heavy

Hence we proposed a partial dynamic reconfiguration to alleviate the execution consumption.

**Authorizing publication:** The MIRACL and the INS laboratory authorize the publication of this work and declare that the material presented isn't classified.

**Acknowledgment:** The authors would thanks Professor Martine Gavaret for all the useful help in the validation of the results of detection of epileptic transient activities.

**Authors signature:**

Nawel Jmail : Sfax University, MIRACL lab , Sfax, Tunisia.
naweljmail@yahoo.fr

Tarek Frikha : Sfax University, CES lab, Sfax, Tunisia.

Abir Hadriche : Gabes University, REGIM, Sfax, Tunisia.
abirhadrich@yahoo.fr

Christian-G Bénar : Aix-Marseille University, Faculty of Medicine, INSERM, Institut de Neurosciences des Systèmes, UMR,1106 Marseille, France.
Christian.benar@univamu.fr